\documentstyle[12pt]{article}

\newcommand{\EQ}{\begin{equation}}
\newcommand{\EN}{\end{equation}}
\begin{document}

\topmargin 0pt
\oddsidemargin=-0.4truecm
\evensidemargin=-0.4truecm
\renewcommand{\thefootnote}{\fnsymbol{footnote}}
\newpage
\setcounter{page}{0}
\begin{titlepage}
\vspace{0.8cm}
\begin{center}
{\large RESONANT SPIN-FLAVOR PRECESSION OF NEUTRINOS AS
A POSSIBLE SOLUTION TO THE SOLAR NEUTRINO PROBLEM
\footnote{Talk given at the XII Moriond Workshop "Massive Neutrinos.
Tests of Fundamental Symmetries", Les Arcs, France, Jan. 25-Feb. 1,
1992}}\\
\vspace{0.4cm}
\begin{flushright}
SISSA Ref. 49/92/EP
\end{flushright}
\vspace{0.4cm}
{\large Eugueni Kh. Akhmedov
\footnote{on leave from Kurchatov Institute of Atomic
Energy, Moscow 123182, Russia}
\footnote{E-mail: akhmedov@tsmi19.sissa.it, ~akhm@jbivn.kiae.su}} \\
\vspace{0.2cm}
{\em Scuola Internazionale Superiore di Studi Avanzati\\
Strada Costiera 11, I-34014 Trieste, Italy} \\
\end{center}
\vspace{1.2cm}

\begin{abstract}
Recent developments of the resonant neutrino spin-flavor precession
scenario and its applications to the solar neutrino problem are reviewed.
We discuss in particular the possibilities of reconciliation
of strong time variations of the solar neutrino flux observed in the
Homestake  ${}^{37}\!$Cl experiment with little or no time variation seen
in the Kamiokande II experiment.
\end{abstract}
\vspace{2cm}
\centerline{March 1992}
\vspace{.5cm}
\end{titlepage}

\renewcommand{\thefootnote}{\arabic{footnote}}
\setcounter{footnote}{0}
\newpage
\section{Introduction}
There are two issues in the solar neutrino problem: \\
{}~(1) the deficiency of solar neutrinos observed in the Homestake
\cite{Davis}, Kamiokande II \cite{Kam} and, most recently, SAGE
\cite{SAGE} experiments;\\
{}~(2) time variation of the solar neutrino flux in anticorrelation with
solar activity (11-yr variations) for which there is a strong indication
in the chlorine experiment of Davis and his collaborators but which is
not seen in the Kamiokande data.

In this talk I will discuss mainly the second issue with the emphasis on
various possibilities of conciliation of the strong time variations
in the Homestake experiment with no or little variation in Kamiokande II.

The most natural explanation of the time variation of the solar neutrino
flux is related to the possible existence of a large magnetic or electric
dipole moments of neutrinos, $\mu \sim 10^{-11}\mu_{B}$. As was pointed
out by Vysotsky, Voloshin and Okun (VVO) \cite{VV,VVO}, strong toroidal
magnetic field in the convective zone of the sun $B_{\bot}$ could then
rotate left-handed electron neutrinos $\nu_{eL}$ into right-handed
$\nu_{eR}$ which escape the detection. In the periods of quiet sun the solar
magnetic field is much weaker and the neutrino spin precession is less
efficient which explains the 11-yr variation of the neutrino flux.

Subsequently, it was noted \cite{VVO,BF} that the matter effects can
suppress the neutrino spin precession. The reason for this is that
$\nu_{eL}$ and $\nu_{eR}$ are not degenerate in matter since $\nu_{eL}$
interact with medium whereas $\nu_{eR}$ are sterile, and their energy
splitting reduces the precession probability. It was also shown
\cite{AKHM1} that, unlike in the MSW effect case, the adiabaticity
may play a bad role for the VVO effect resulting in a reflip of neutrino
spin and thus reducing the probability of $\nu_{eL} \rightarrow \nu_{eR}$
transition. In order to break the adiabaticity, the precession length
should be large as compared to the characteristic lengths over which
matter density and magnetic field vary significantly, which gives an upper
bound on $\mu B_{\bot}$. This parameter should be  also bounded from below
in order for the precession phase not to be too small. Therefore one gets a
rather narrow range of allowed values of $\mu B_{\bot}$ \cite{AKHM1}.

Another interesting possibility is the neutrino spin-flavor precession
(SFP) due to the interaction of flavor-off-diagonal (transition) magnetic
or electric dipole moments of neutrinos $\mu_{ij}$ with transverse magnetic
fields \cite{ShVa,VVO}. The SFP is the rotation of neutrino spin with its
flavor being simultaneously changed. Such a process can occur even for
Majorana neutrinos since the $CPT$ invariance does not preclude the
transition magnetic dipole moments of Majorana particles. Until recently,
the neutrino SFP has not attracted much attention because it was expected
to be suppressed by the energy splitting of the neutrinos of different species.
If the "Zeeman energy" $\mu_{ij} B_{\bot}$ is small as compared to the
kinetic energy difference $\Delta m^{2}_{ij}/2E$, the SFP probability is
heavily suppressed. However, in 1988 it was noted independently by the
present author \cite{AKHM2,AKHM1} and by Lim and Marciano \cite{LM} that
in matter the situation can change drastically. Since $\nu_{eL}$ and
right-handed neutrinos or antineutrinos of another flavor interact with
matter differently, the difference of their potential energies can cancel
their kinetic energy difference resulting in a resonant amplification of
the SFP. Therefore in matter the SFP of neutrinos can be enhanced, unlike
the VVO neutrino spin precession
\footnote{The VVO neutrino spin rotation can also be resonantly enhanced
provided the magnetic field twists along the neutrino trajectory, see
\cite{SM,AKS} and below.}. The resonant spin-flavor precession (RSFP)
of neutrinos has also some more advantages as compared to the VVO mechanism:

$\bullet$ the adiabaticity plays a good role for the RSFP increasing the
conversion
probability, and therefore the $\mu_{ij}B_{\bot}$ should be bounded only
from below; the required magnitude of this parameter is a factor of
$2-3$ smaller than that for the VVO effect;

$\bullet$ some energy dependence of the neutrino conversion seems to be
necessary to reconcile the Homestake and Kamiokande II data (see below).
The RSFP probability has the desired energy dependence whereas
the VVO neutrino spin precession is energy independent.

Although the above arguments disfavor the VVO effect as a solution of the
solar neutrino problem, they do not rule it out, given the
uncertainties of the experimental data.
\section{General features of RSFP of neutrinos}
The RSFP of neutrinos is analogous to the resonant neutrino oscillations
\cite{MS,W}, but differs
from the latter in a number of important respects. The main features of
this effect have been discussed in detail in my talk at the last Moriond
meeting \cite{AKHM3}, and so I will just briefly mention them here.

The magnetic-field induced mixing of $\nu_{eL}$ and $\nu_{\mu R}
(\bar{\nu}_{\mu R})$ can be described by the mixing angle $\theta$,
\EQ
\tan 2\theta = \frac{2\mu_{e\mu}B_{\bot}}{\sqrt{2}G_{F}(N_{e}-
\alpha N_{n})-\frac{\Delta m^{2}_{e\mu}}{2E}\cos 2\theta_{0}}
\EN
Here $N_{e}$ and $N_{n}$ are the electron and neutron number densities,
$\alpha=1/2$ for Dirac neutrinos and 1 for Majorana neutrinos, $G_{F}$
is the Fermi constant, and $\theta_{0}$ is the ordinary neutrino mixing
angle in vacuum. The resonant density is defined as a density at
which the mixing angle $\theta$ becomes $\pi/4$:
\EQ
\sqrt{2}G_{F}(N_{e}-\alpha N_{n})|_{r}=\frac{\Delta m^{2}_{e\mu}}{2E}
\cos 2\theta_{0}
\EN
The efficiency of the $\nu_{eL}$$\rightarrow$$\nu_{\mu R}
(\bar{\nu}_{\mu R})$ transition is defined by the degree of the adiabaticity
which depends on both the neutrino energy and magnetic field strength at the
resonance:
\EQ
\lambda\equiv \pi \frac{\Delta r}{l_{r}}=
8\frac{E}{\Delta m^{2}_{e\mu}}(\mu_{e\mu}B_{\bot r})^{2}
L_{\rho}
\EN
Here
\EQ
\Delta r=\frac{8E\mu_{e\mu}B_{\bot r}}{\Delta m^{2}_{e\mu}}L_{\rho}
\EN
is the resonance width, $l_{r}=\pi/\mu_{e\mu}B_{\bot}$ is the precession
length at the resonance and $L_{\rho}$ is the characteristic length over
which matter density varies significantly in the sun.
For the RSFP to be efficient, $\lambda$ should be $> 1$. In
non-uniform magnetic field the field strength at resonance $B_{\bot r}$
depends on the resonance coordinate and so, through eq. (2), on
neutrino energy. Therefore the energy dependence of the adiabaticity
parameter $\lambda$ in eq. (3) is, in general, more complicated than
just $\lambda\sim E$, and is defined by the magnetic field profile
inside the sun
\footnote{Note that for the MSW effect the adiabaticity parameter
is inversely proportional to $E$ \cite{MS}.}.
The main difficulty in the analyses of the RSFP as a possible solution
of the solar neutrino problem is that this profile is essentially
unknown, so that one is forced to use various more or less plausible
magnetic field configurations.

In the adiabatic regime $(\lambda >>1)$, the $\nu_{eL}$ survival
probability is
\EQ
P(\nu_{eL}\rightarrow \nu_{eL})=\frac{1}{2}+\frac{1}{2}\cos 2\theta
_{i}\cos 2\theta_{f}+\frac{1}{2}\sin 2\theta_{i}\sin 2\theta_{f}
\cos \int\nolimits_{t_{i}}^{t_{f}}\Delta E(t)\,dt
\EN
where
\EQ
\Delta E=\sqrt{\left[\sqrt{2}G_{F}(N_{e}-\alpha N_{n})-
\frac{\Delta m^{2}_{e\mu}}{2E}\cos 2\theta_{0}\right]^{2}+(2\mu_{e\mu}
B_{\bot})^{2}}
\EN
Here $\theta_{i}$ and $\theta_{f}$ are the mixing angles (1) at the
neutrino production point and on the surface of the sun respectively.
If the $\nu_{eL}$ are produced at a density which is much higher
than the resonant one, $\theta_{i}\approx 0$ and the survival
probability (4) becomes
\EQ
P(\nu_{eL}\rightarrow \nu_{eL})\approx \cos^{2} \theta_{f}
\EN
Since the magnetic field becomes very weak at the sun's surface, the
mixing angle $\theta_{f}\approx \pi/2$, and so the $\nu_{eL}$ survival
probability is very small in the adiabatic regime. The adiabaticity
parameter $\lambda$ in eq. (3) depends
drastically on the magnetic field strength at resonance, which gives a natural
explanation of time variations of the solar magnetic flux in
anticorrelation with solar activity.

The RSFP requires non-vanishing flavor-off-diagonal magnetic dipole
moments of neutrinos and so is only possible if the neutrino
flavor is not conserved. Therefore neutrino oscillations must also
take place, and in general one should consider the SFP and oscillations
of neutrinos jointly. This have been done in a number of papers both
analytically \cite{AKHM4,AKHM5} and numerically
\cite{LM,AKHM4,MN,BHL,AKHM5}. It was shown that a subtle interplay
between the RSFP and the
MSW resonant neutrino oscillations can occur. In particular,  although
the resonant neutrino oscillations cannot give rise to the time variations
of the solar neutrino flux, they can assist the RSFP to do so by improving the
adiabaticity of the latter \cite{AKHM5}.
\section{Neutrino spin precession in twisting magnetic fields}
If the magnetic field changes its direction along the neutrino trajectory,
this can result in new interesting phenomena. In particular, new kinds of
resonant neutrino conversions become possible, the energy
dependence of the conversion probability can be significantly distorted
and the lower limit on the value of $\mu B_{\bot}$ required to account
for the solar neutrino problem can be slightly relaxed \cite{SM,AKS}.
Moreover, if the neutrino oscillations are also taken into account,
the transitions $\nu_{e}\rightarrow \bar{\nu}_{e}$ can become resonant, and
the order of the RSFP and MSW resonances can be interchanged \cite{AKMSP}.

Since the main features of the resonant neutrino spin-flip transitions in
twisting magnetic fields are discussed in some detail in the contributions
of Krastev and Toshev in this volume, I will confine myself to a new
development which was not covered in their talks.

A few years ago, Vidal and Wudka \cite{VW} claimed that the field rotation
effects can greatly enhance the neutrino spin-flip probability and reduce the
needed value of $\mu B_{\bot}$ by a few orders of magnitudes. In
\cite{SM,AKS} it was shown that this result is incorrect and typically
the required
value of $\mu B_{\bot}$ can only be reduced by a factor 2--3 (see also
\cite{ASh1,ASh2} in which the process without matter effects was considered).
However, in these papers it was not proved that there cannot exist a rotating
field configuration giving stronger enhancement of the spin-flip
probability and larger gain in the $\mu B_{\bot}$ parameter. Recently, Moretti
\cite{M} has found a severe constraint on the transition probability which
eliminates even this possibility. The effective
Hamiltonian describing the evolution of the system of left handed $\nu_{eL}$
and right handed neutrino of the same or another flavor $\nu_{R}$ in a
twisting magnetic field is
\EQ
H=\left(\begin{array}{cc}
   V(t)/2 & \mu B_{\bot}e^{i\phi (t)}\\
   \mu B_{\bot}e^{-i\phi (t)} & -V(t)/2
\end{array}
\right)
\EN
where V(t) is just the denominator of the r.h.s. of eq. (1), and the angle
$\phi (t)$ defines the direction of the
magnetic field in the plane orthogonal to the neutrino momentum. The
transition probability $P(\nu_{eL}\rightarrow \nu_{R})$ turns out to have
the following upper bound \cite{M}:
\EQ
P(\nu_{eL}\rightarrow \nu_{R};t)\leq \mu\int\nolimits_{0}^{t}B_{\bot}
(t^{'})\,dt^{'}
\EN
The analogous result can also be obtained for the  neutrino oscillations
in matter as well as for the evolution of any other two-level system.
\section{RSFP and antineutrinos from the sun}
If both the SFP and oscillations of neutrinos can occur, this will result
in the conversion of a fraction of solar $\nu_{e}$ into $\bar{\nu}_{e}$
\cite{LM,AKHM4,AKHM6,RBL}. For Majorana neutrinos, the direct $\nu_{e}
\rightarrow \bar{\nu}_{e}$ conversions are forbidden since the $CPT$
invariance precludes the diagonal magnetic moment $\mu_{ee}$. However,
this conversion can proceed as a two-step process in either of two ways:
\begin{eqnarray}
\nu_{eL}\stackrel{\rm oscill.}{\longrightarrow}\nu_{\mu L}\stackrel
{\rm SFP}{\longrightarrow}\bar{\nu}_{eR}\\
\nu_{eL}\stackrel{\rm SFP}{\longrightarrow}\bar{\nu}_{\mu R}\stackrel
{\rm oscill.}{\longrightarrow}\bar{\nu}_{eR}
\end{eqnarray}
One can then consider two possibilities:

{}~(1) both oscillations and SFP take place inside sun \cite{LM,AKHM4,AKHM6}.
The amplitudes of the processes (10) and (11) have opposite signs
since the matrix of the magnetic moments of Majorana neutrinos is
antisymmetric. Therefore there is a large cancellation between these two
amplitudes (the cancellation is exact in the limit of vanishing neutron
density $N_{n}$), and the probability of the $\nu_{e}\rightarrow
\bar{\nu}_{e}$ conversion inside the sun turns out to be about 3--5\%
even for large mixing angles $\theta_{0}$ \cite{AKHM4,AKHM6}.

{}~(2) Only the RSFP transition $\nu_{eL}\rightarrow\bar{\nu}_{\mu R}$
occurs in the sun with an appreciable probability whereas the oscillations
of neutrinos proceed mainly in vacuum on their way between the sun and the
earth [eq. (11)]. For not too small neutrino mixing
angles the probability of the $\nu_{e}\rightarrow \bar{\nu}_{e}$ conversion
can then be quite sizable \cite{RBL}.

In \cite{BFMM} the background events in the Kamiokande II experiment were
analysed and a stringent bound on the flux of $\bar{\nu}_{e}$ from the sun
was obtained: $\Phi(\bar{\nu}_{e})\leq (0.05-0.07)\Phi(\nu_{e})$.
This poses a limit on the models in which both the RSFP and neutrino
oscillations occur: the mixing angle $\theta_{0}$ should be less than
$6-8^{\circ}$. This rules out the models with the large magnetic moments
of pseudo Dirac neutrinos including those with only one neutrino
generation for which $\theta_{0}$ is the mixing between $\nu_{eL}$ and
sterile $\bar{\nu}_{eL}$ \cite{KLN,MN2}. However, the models with a
conserved lepton charges $L_{e}\pm (L_{\mu}-L_{\tau})$ are not
excluded even though the mixing angle is $\pi/4$, since the $\nu_{e}
\rightarrow \bar{\nu}_{e}$ conversion probability vanishes identically
in this case \cite{BFMP}.

The $\bar{\nu}_{e}$ production due to the combined effect of
the RSFP and oscillations of neutrinos can be easily distinguished from
the other mechanisms of $\bar{\nu}_{e}$ generation (like $\nu\rightarrow
\bar{\nu}+{\rm Majoron}$ decay) since (i) the neutrino flux should vary
in time in {\em direct} correlation with solar activity, and (ii) the
neutrino energy is not degraded in this case \cite{AKHM4,AKHM6}. The
$\bar{\nu}_{e}$ flux from the sun of the order of a few per cent of
the expected $\nu_{e}$ flux should be detectable in the
forthcoming solar neutrino experiments like BOREXINO, SNO and
Super-Kamiokande \cite{AKHM6,RBL,BL}.
\section{Reconciling the Homestake and Kamiokande II data}
It has been mentioned above that while there is a strong indication
in favor of time variation of the neutrino detection rate in the Homestake
data, the Kamiokande experiment does not see such a time variation.
It still cannot rule out a small $(\leq 30\%)$ time variation.
Therefore a question naturally arises as to whether it is possible to
reconcile large time variations in the Homestake ${}^{37}\!$Cl experiment
with small time variation in the water \v{C}erenkov experiment. There are
two major differences between these two experiments which could in principle
give rise to different time variations of their detection rates:

{}~(1) Homestake experiment utilizes the $\nu_{e}-{}^{37}\!$Cl charged current
reaction, while in the Kamiokande detector $\nu -e$ scattering is used
which is mediated by both charged and neutral currents;

{}~(2) the energy threshold  in the Homestake experiment is 0.814 MeV so that
it is sensitive to high energy ${}^{8}$B, intermediate energy
${}^{7}$Be and partly to low energy $pep$ neutrinos; at the same time the
energy threshold  in the Kamiokande II experiment is 7.5 MeV and so it is
only sensitive to the high-energy fraction of the ${}^{8}$B neutrinos.

In \cite{AKHM7,AKHM5} it was noted that if the lower-energy neutrino
contributions to the chlorine detection rate are suppressed stronger than
that of high-energy neutrinos, the latter can vary in time with smaller
amplitude and still fit the Homestake data. In that case one can expect
weaker time variations in the Kamiokande II experiment. The desired
suppression of the low-energy neutrino flux can be easily explained in
the framework of the RSFP scenario as a consequence of flavor-changing
spin-flip conversion due to a strong inner magnetic field, the existence
of which seems quite plausible \cite{C}. The alternative possibility is
the suppression of low energy neutrinos by the MSW effect when RSFP and
the resonant neutrino oscillations operate jointly. Another important
point is that due to the RSFP solar $\nu_e$ are converted into
$\bar{\nu}_{\mu R}$ or $\bar{\nu}_{\tau R}$ which are sterile for the
chlorine detector but can be detected (though with a smaller cross
section) by water \v{C}erenkov detectors. This also reduces the
amplitude of the time variation in the Kamiokande II detector. If both
these factors are taken into account, it becomes possible to
reconcile the Homestake and Kamiokande data; one can expect a low
signal in the gallium experiments in this case since they are primarily
sensitive to low energy neutrinos whose flux is supposed to be heavily
suppressed \cite{AKHM7,AKHM5}.

A similar possibility has been recently considered by
Babu, Mohapatra and Rothstein \cite{BMR} and by Ono and Suematsu
\cite{OS}. They pointed out that due to the energy dependence of the
RSFP neutrino conversion probability, lower-energy neutrinos can exhibit
stronger time variations (i.e. stronger magnetic field dependence)
than the higher-energy ones. In fact, this is very
natural in the RSFP scenario: with increasing neutrino energy the width
of the resonance increases [see eq. (4)] and at sufficiently high energies
it can be a significant fraction of the solar radius. The neutrino
production point can then happen to be inside the resonant region, which
reduces the conversion efficiency. The different magnetic field dependence
of the Homestake and Kamiokande II detection rates is illustrated by the
figures which we borrowed from ref. \cite{BMR}.

\vspace*{7.5truecm}
\noindent
{\footnotesize
\vspace*{-.2truecm}
Fig. 1. (a) Expected event rate in chlorine as a function of the convective
zone magnetic field. Here
\vspace*{-.2truecm}
$\Delta m^{2}=7.8\times 10^{-9}$
eV$^{2}$, the maximal value of the magnetic field in the core $B_{1}=10^{7}~G$
and $\mu =2\times 10^{-11}\mu_{B}$.
\vspace*{-.2truecm}
(b) The same as (a) but for the Kamiokande event rate.}\\

It should be noted that the
ordinary VVO neutrino spin precession lacks energy dependence which is
required to get smaller time variation in the Kamiokande II experiment.
Moreover, it converts $\nu_{eL}$ into sterile $\nu_{eR}$ (unless
the neutrinos are Zeldovich-Konopinski-Mahmoud particles) which do not
contribute to the $\nu -e$ cross section. However, for the VVO scenario yet
another possibility of reconciliation of the Homestake and Kamiokande data
exists. In order to get sizable magnetic moments of neutrinos, $\mu\approx
10^{-11}\mu_{B}$, one has to go beyond the Standard Model. Most of the
models producing large neutrino magnetic moments are based on various
extensions of the Standard Model containing new charged scalars.
In these models right-handed sterile neutrinos can interact with electrons
via scalar exchange and therefore can contribute
to the $\nu -e$ reaction which increases the signal in the Kamiokande II
detector and reduces the amplitude of its time variation \cite{FY}.
Note that the models giving large transition neutrino magnetic moments
usually also contain new scalars and therefore the same mechanism
can be operative in case of the RSFP as well.
\section{Conclusion}
We conclude that the resonant neutrino spin-flavor precession mechanism
provides a viable explanation of the solar-neutrino problem which complies
with all the existing experimental data and yields a number of interesting
predictions for the forthcoming experiments.
\section*{Acknowledgement}
The author is deeply indebted to
Scuola Internazionale Superiore di Studi Avanzati where this report was
written for kind hospitality and financial support.

\end{document}